# Properties of Ferromagnetic $Ga_{1-x}Mn_xN$ Films Grown by Ammonia-MBE

Saki Sonoda, Hidenobu Hori, Yoshiyuki Yamamoto, Takahiko Sasaki, Masugu Sato, Saburo Shimizu, Ken-ichi Suga and Koichi Kindo

*Abstract*—Using ammonia as nitrogen source for molecular beam epitaxy, the GaN-based diluted magnetic semiconductor $Ga_{1-x}Mn_xN$ is successfully grown with Mn concentration up to x~6.8% and with p-type conductivity. The films have wurtzite structure with substitutional Mn on Ga site in GaN. Magnetization measurements revealed that $Ga_{1-x}Mn_xN$ is ferromagnetic at temperatures higher than room temperature. The ferromagnetic-paramagnetic transition temperature, Tc, depends on the Mn concentration of the film. At low temperatures, the magnetization increases with increasing of magnetic field, implying that a paramagnetic-like phase coexists with ferromagnetic one. Possible explanations will be proposed for the coexistence of two magnetic phases in the grown films.

## I. INTRODUCTION

Diluted magnetic semiconductors (DMSs) have created much interest in recent years since the first fabrication of Mn doped InAs, (In,Mn)As [1] and the discovery of ferromagnetism in (Ga,Mn)As [2] . Especially the system of (Ga,Mn)As is studied widely, resulting in the successful fabrication of spin electronic devices such as spin light emitting diode (LED) [3]. However, in this system, there is a serious problem, their Curie temperature is as low as 110K observed for $Ga_{0.947}Mn_{0.053}As$ [4], [5]. To understand the ferromagnetism in these semiconductors, and to find answers for overcoming the problem, some theoretical studies have been carried out and expectations of room temperature ferromagnetism in wide-gap semiconductors have been reported [6], [7]. These theoretical predictions have been inspired many efforts to realize DMSs based on wide-gap materials [8]-[16].

Wurtzite GaN, one of the most famous wide-gap semiconductors, is a promising material for optical and electronic devices. In fact, the recent development of growth techniques for wurtzite III-nitrides has resulted in the successful fabrication of GaN-based optical and electronic devices [17], [18]. Hence it can readily be imagined that the realization of ferromagnetism with high Tc in GaN film is of exceeding practical significance in considering their high potential as materials for spintronic devices.

According to the theoretical predictions on DMSs, one has to realize p-type GaN doped with a transition metal in order to make the material a ferromagnet, while GaN is notorious for the difficulty to realize p-type conduction in the film, especially for MBE method. However, considering future application of the films that might need monolayer-order controlled interfaces, MBE growth is considered to be indispensable because of its well-known advantages, which are significantly reduced growth temperatures resulting in decreased residual strain and improved interfaces and doping profiles due to reduced diffusion. For this reason, in spite of the difficulties in MBE growth of III-nitride films with p-type conductivity, many efforts have been made to obtain p-type GaN for optical devices, then resulted in successful growth of p-type GaN:Mg using ammonia as nitrogen source. By this technique, so called ammonia-MBE, one can obtain p-type GaN films without after-growth thermal annealing [22], [23].

Recently, we have succeeded in the growth of ferromagnetic Mn-doped GaN films with Tc of 940K by ammonia-MBE [24]-[26]. In this paper, we report on the magnetic and the electrical properties of the ammonia-MBE grown $Ga_{1-x}Mn_xN$ films having various Mn concentrations with emphasis on the crystallographic properties of the films.

## II. EXPERIMENTAL

The growth of $Ga_{1-x}Mn_xN$ films was carried out in an ammonia-MBE system (ULVAC, MBC-100) equipped with a reflection high-energy electron diffraction (RHEED) apparatus. Solid source effusion cells were used as Ga and Mn sources while $NH_3$ gas was used as nitrogen source. No intentional doping except Mn was used throughout the present work. Four samples ware prepared for this study. The substrate temperature for the Mn-doped GaN films was 720°C. The film thicknesses of the three samples are 3600Å (sample A), 3000Å (sample B), 4700Å (sample C) and 4700 Å (sample D). The

Manuscript received February 1, 2002. This work was supported in part by the Ministry of Education, Culture, Sports, Science and Technology, Government of Japan (MONBUKAGAKUSHO).

S. Sonoda (e-mail: saki_sonoda@ulvac.com) and S. Shimizu (e-mail: saburo_shimizu@ulvac.com) are with ULVAC Inc, 2500 Hagisono, Chigasaki, Kanagawa, 253-8543 Japan (telephone: +81-463-89-2057).
H. Hori (e-mail: h-hori@jaist.ac.jp), Y. Yamamoto (e-mail: y-yamamo@jaist.ac.jp) and T. Sasaki (e-mail: t-sasaki@jaist.ac.jp) are with School of Materials Science, Japan Advanced Institute of Science and Technology (JAIST), 1-1, Asahidai, Tatsunokuti, Ishikawa, 923-1292 Japan.
M. Sato (e-mail: msato@spring8.or.jp) is with Japan Synchrotron Radiation Research Institute (JASRI), Mikazuki, Hyogo 679-5198, Japan.
K. Suga (e-mail: suga@mag.rcem.osaka-u.ac.jp) and K. Kindo (e-mail: kindo@rcem.osaka-u.ac.jp) are with Research Center for Materials Science at Extream Conditions (KYOKUGEN), Osaka University, 1-3 Machikanayama, Toyonaka, Osaka 560-8531, Japan.

Mn concentrations of the films were estimated with the aid of an electron probe micro analyser. The Mn concentration x of the $Ga_{1-x}Mn_xN$ films of sample A, sample B, sample C and sample D are 0.057, 0.068, 0.029 and 0.070, respectively. No after-growth thermal treatment was carried out for all samples.

### III. RESULTS AND DISCUSSION

#### A. Crystallographic Properties

As the crystallographic assumption of GaN-based DMS, it must be confirmed that Mn is located on Ga site as substitutional atom in the films with neither phase nor surface segregation of a ferromagnetic phase.

*1) Local Lattice Configuration*

Extended X-ray absorption fine structure (EXAFS) analyses were carried out to study the local structure of Mn atoms in the grown films. The details are given in ref. [27], only the outline is described below. The shape of the Fourier transform of Mn K-edge EXAFS spectra of sample A is very similar with that of the Ga K-edge of sample A. Also the both experimental results on Mn and Ga K-edges correspond well to the calculated ones for wurtzite structure. These results mean that the local structure of Mn corresponds well to that of Ga in wurtzite GaN. Hence, it can be stated that the majority of the Mn atoms are substituting the Ga in the GaN.

*2) Surface Characteristics*

The GaMnN films of sample A [24]-[26], sample B and sample D produce a streak RHEED pattern during growth, whilst non-doped GaN and GaMnN of sample B spotty patterns. This implies that Mn is a good catalyst for decomposition of $NH_3$ into nitrogen and hydrogen same as Mg [28]. As for the structure of subsurface region of the $Ga_{1-x}Mn_xN$ film, it has already confirmed and reported that it has the wurtzite structure with a depth of 10Å by coaxial impact collision ion scattering spectroscopy (CAICISS) [29]-[31]. Also it was shown that there is no trace of surface segregation of manganese by CAICISS [31].

*3) Crystal Structure and Lattice Constants*

The crystal structure of the 4 samples was confirmed as wurtzite structure by X-ray diffraction measurements (Philips, X'Pert MRD). No peaks of other phases were observed. Because the epitaxal films are wurtzite, they have 2 lattice constants *a* and *c*. It should be noted that the growth direction of $Ga_{1-x}Mn_xN$ and GaN buffer layers are parallel to the c-axis of the wurtzite structure. In Fig. 1, the lattice constants *a* and *c* in the GaN buffer layer, sample C and sample D are plotted as the Mn concentration dependences. It revealed that the lattice constant *a* of the as-grown wurtzite $Ga_{1-x}Mn_xN$ film is larger than that of the GaN buffer layer, while the lattice constant *c* of the wurtzite $Ga_{1-x}Mn_xN$ film is a bit smaller than that of the GaN buffer layer. Also it was found that the films with higher Mn concentration have larger *a* and shorter *c*. To investigate the effect of Mn-doping on crystalline quality, we have been carring out XRD measurements across various Bragg reflections. Figure 2 shows the Mn concentration dependences of the full width at half maximum (FWHM) of the GaN buffer layer, sample C and sample D. The simple decreases of all the FWHMs with the increase of Mn concentration indicate that the film with higher Mn concentration has higher crystalline perfection. This is consistent with the results of the observation of RHEED patterns.

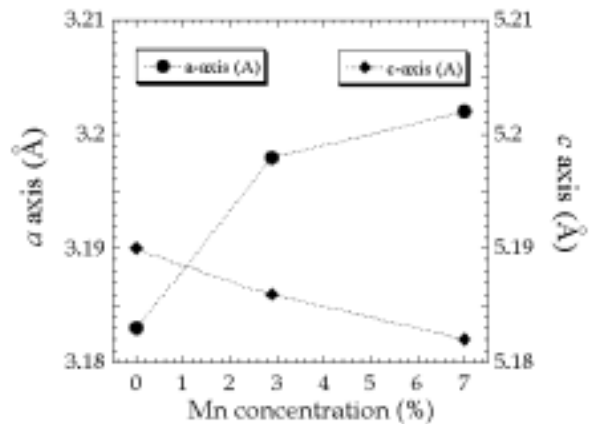

Fig. 1. Mn concentration dependences of the lattice constants *a* and *c*.

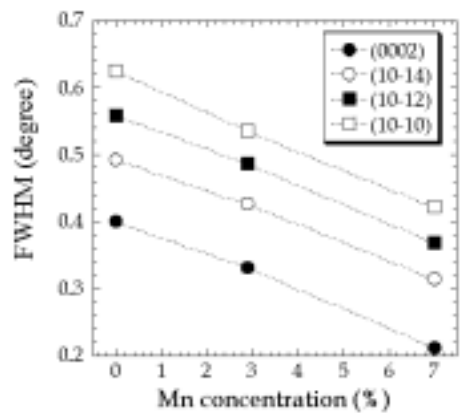

Fig. 2. Mn concentration dependences of the FWHMs of XRD across Bragg reflections (0002), (10-14), (10-12) and (10-10).

*4) Colour of The Films*

The colour of the samples is light reddish-brown yet transparent and the colour becomes lighter with a decrease of Mn concentration of $Ga_{1-x}Mn_xN$ layer. These facts can support the suggestions that Mn is incorporated into the GaN structure not in segregated metal phase.

#### B. Magnetic and Electrical Properties

Magnetization measurements of the films were carried out in a superconducting quantum interference device (SQUID) - magnetometer in the temperature range from 2.0 to 400 K. Figure 3 shows the *M-H* curve of sample A in field up to 0.2T at the temperatures 2K, 4.2K, 10K, 100K and 400K. Contributions from the sapphire substrate and the GaN buffer layer were not subtracted in Fig 3. From these M-H curves, it can be stated that the $Ga_{0.943}Mn_{0.057}N$ film of sample A is ferromagnetic at 300K and 400K.

We have carried out two kinds of discussions on the contribution of Mn atoms to ferromagnetic behaviour: (I) the way after the example of (Ga,Mn)As [32]-[34] and (II) the calculation based on our assumption [35]. (I) Magnetic moments per Mn atom at 4.2K and 300K were estimated as $5.3\mu_B$ and $1.1\mu_B$, respectively, with the magnetization values of

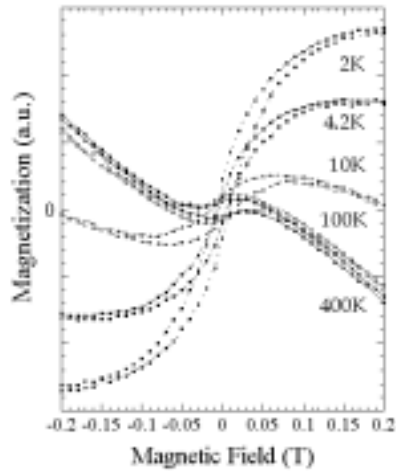

Fig. 3  *M-H* curves of sample A in fields up to 0.2T at the temperatures 2K, 4.2K, 10K, 100K and 400K.

sample A at 7T [25], [26]. Assuming divalent Mn, about 20% of the Mn in the film contributes to the ferromagnetic behaviour. (II) We should note that the steep increase of the magnetization was observed at the low temperature, which could not be expected for the ferromagnetic moment with high Tc. This behaviour could be explained by assuming that the magnetization consists of temperature independent ferromagnetic moment and temperature dependent paramagnetic one. At room temperature, paramagnetic moment is suppressed by thermal fluctuation resulting in that the ferromagentic behaviour is dominant while the paramagnetic moment recovers with decreasing temperature in accordance with the Curie law. Calculation [35] based on the above assumption gives a good agreement with the experimental results. It is also found from the calculation that about a half of the Mn ions contributes to the paramagnetic curve and the rest to the ferromagnetic one, and the magnetic moment per the Mn atom is found to be about $4\mu_B$ and $1\mu_B$ for

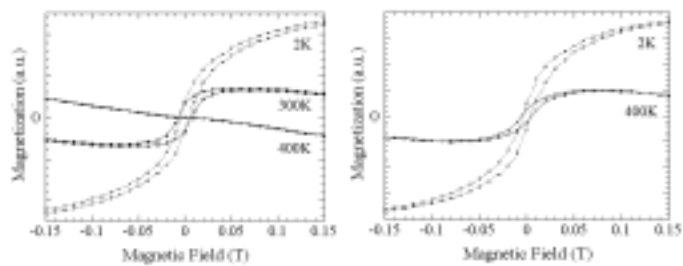

Fig. 4  (a) *M-H* curves of sample B at the temperatures 2K and 400K, (b) *M-H* curves of sample C at the temperatures 2K, 300K and 400K.

paramagnetic and ferromagnetic moment, respectively. Thus, the results on the magnetization curves can be understood by considering coexisting para- and ferromagnetic phases though the origin of the coexistence is not clear.

M-H curves of sample B and sample C are shown in Figs. 4(a) and 4(b). What has to be noticed, is the difference between the hysteresis loops at 400K of these samples. Sample B with the Mn concentration x=0.068 shows obvious hysteretic behaviour indicating spontaneous magnetization even at 400K which is the same as the sample A. On the other hand, the hysteresis loop of sample C with Mn concentration x=0.029 is almost flat, i.e. critical temperature is considered to be slightly higher than 400K. We have not carried out magnetization measurements in the high temperature region for these samples yet. However, from the temperature dependences of these M-H curves, it can be stated that the Curie temperatures depends on the Mn concentration of the GaMnN films.

Regarding the electrical properties, the carrier type of the Mn-doped GaN films were estimated as p, while the GaN buffer layer showed n-type conductivity [25].

To investigate the coexistence of the ferromagnetic phase and the paramagnetic-like in more detail, we have been carrying out X-band electron spin resonance (ESR) measurements on the samples at room temperature. As details of the ESR measurements will be discussed in [36], only a brief outline and representative data are given in this paper. Figure 6 shows the ESR spectra of sample B. Each spectrum,

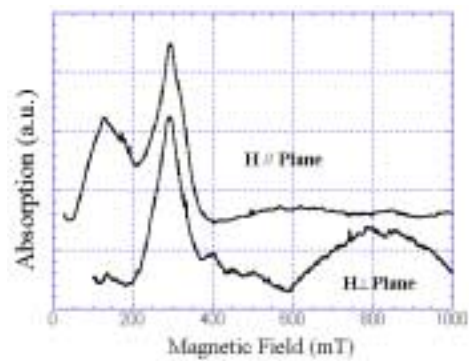

Fig. 5   ESR spectra of sample B.

that is the spectrum from the sample placed in magnetic field parallel to the film and that in magnetic field perpendicular to the film, displays two peaks: a sharp peak at around 290mT and a broader peak around 125mT and 810mT for B// and B⊥, respectively. From the fact that the sharp peak does not depend on the direction of the magnetic field, it can be stated that this peak is derived from the paramagnetic phase. On the other hand, the broader peak depends on the direction: when the magnetic field is applied perpendicular to the film, the absorption occurs at higher magnetic field. Assuming that this shift is mainly due to the demagnetization field, this resonance is derived from the ferromagnetic phase. These results mean that the coexistence of two magnetic phases is confirmed using

ESR measurements. Quantitative analyses in g-values with detailed discussion will be given in [36].

The temperature dependence of the ESR and the calculations for the $Ga_{1-x}Mn_xN$ film with various Mn concentrations will give us more information to understand the behaviour in the magnetization.

## IV. Conclusion

GaMnN films grown by ammonia-MBE were investigated. The films have wurtzite structure with substitutional Mn on the Ga site in GaN without phase segregation. The films with higher Mn concentrations have higher crystallographic perfection. The films show ferromagnetic behaviour with p-type conductivity. The Curie temperature is controllable with Mn concentration up to x=6.8%. At low temperatures, the magnetization does not saturate in fields of 0.2. This fact implies that a paramagnetic-like phase coexists with the ferromagnetic one. This coexistence is also observed in electron spin resonance spectra.

## Acknowledgment

The authors would like to thank Mr. T. Okuno, Mr. T. Fujino, Mr. S. Shindo, Associate Prof. Dr. M. Katayama and Prof. Dr. K. Oura of Osaka University for their study on the material by CAICISS. We also thank Dr. T. Kitada, Dr. S. Shimomura and Prof. S. Hiyamizu of Osaka University for analysing XRD, useful discussion and encouragements.

## References

[1] H. Munekata, H. Ohno, S. von Molnar, A. Segmullerr, L. L. Chang and L. Esaki: Phys. Rev. Lett. vol. 63 pp. 1849, 1989.
[2] H. Ohno, H. Munekata, T. Penney, S. von Molnar and L. L. Chang: Phys. Rev. Lett. vol. 68, pp. 2664, 1992.
[3] Y. Ohno, D. K. Young, B. Beschoten, F. Matsukura, H. Ohno and D. D. Awschalom: Nature vol. 402, pp. 709, 1999.
[4] H. Ohno, A. Shen, F. Matsukura, A. Oiwa, A. Endo, S. Katsumoto, and Y. Iye, Appl. Phys. Lett. vol. 69, pp.363, 1996.
[5] F. Matsukura, H. Ohno, A. Shen, and Y. Sugawara, Phys. Rev. B vol. 57, pp. R2037, 1998.
[6] T. Dietl, H. Ohno, F. Matsukura, J. Cibert and D. Ferrand: Science vol. 287, pp. 1019, 2000.
[7] K. Sato and H. Katayama-Yoshida: Jpn. J. Appl. Phys. vol. 40 pp. L485, 2001.
[8] X. A. Cao, S. J. Pearton, G. T. Dang, A. P. Zhang, F. Ren, R. G. Wilson, and J. M. Van Hovee, J. Appl. Phys. vol. 87, pp. 1091, 2000.
[9] H. Akinaga, S. Nemeth, J. De Boeck, L. Nistor, H. Bender, G. Borghs, H. Ofuchi and M. Oshima, Appl. Phys. Lett. vol. 77 pp. 4377, 2000.
[10] H. Ofuchi, M. Oshima, M. Tabuchi, Y. Takeda, H. Akinaga, S. Nemeth, J. De Boeck, and G. Borghs, Appl. Phys. Lett. Vol. 78, pp. 2470, 2001.
[11] M. Zajac, R. Doradzinski, J. Gosk, J. Szczytko, M. Lefeld-Sosnowaka, M. Kaminska, A. Twardowski, M. Palczewska, E. Grzanka and W. Gebicki, Appl. Phys. Lett. vol. 78, pp. 1276, 2001.
[12] N. Theodoropoulou, A. F. Hebard, M. E. Overberg, C. R. Abernathy, S. J. Pearton, S. N. G. Chu and R. G. Wilson, Appl. Phys. Lett. vol. 78, pp. 3475, 2001.
[13] K. Ueda, H. Tabata and T. Kawai, Appl. Phys. Lett., vol. 79, 988, 2001.
[14] Y. Matsumoto, M. Murakami, T. Shono, T. Hasegawa, T. Fukumura, M. Kawasaki, P. Ahmet, T. Chikyow, S. Koshihara and H. Koinuma, Science, vol. 291, 854, 2001
[15] S. Kuwabara, T. Konodo, T. Chikyow, P. Ahmet, and H. Munekata, Jpn. J. Appl. Phys. vol. 40, pp. L724, 2001.
[16] Appl. Phys. Lett.
[17] H. Amano, M. Kito, K. Hiramatsu, and I. Akasaki: Jpn. J. Appl. Phys. vol. 28, pp. L2112, 1989.
[18] H. Morkoc, S. Strites, G. B. Gao, M. E. Lin, B. Sverdlov, and M. Burns: J. Appl. Phys. vol. 76, pp. 1363, 1994.
[19] C. R. Abernathy, GaN and related materials, New York: Gordon and Breach, 1997, pp. 11-51.
[20] N Grandjean, and J. Massies, Appl. Phys. Lett. vol. 71, 1816, 1997.
[21] M. Mesrine, N. Grandjean and J. Massies, Appl. Phys. Lett. vol. 72, 350, 1998.
[22] M. Kamp, M. Mayer, A. Pelzmann, and K. J. Ebeling, Internet Journal of Nitride Semiconductor Research, vol. 2, 26, 1997.
[23] N Grandjean, and J. Massies, Appl. Phys. Lett. vol. 72, 82, 1998.
[24] S. Sonoda, S. Shimizu, T. Sasaki, Y. Yamamoto, and H. Hori, submitted for publication to J. Crystal Growth, cond-mat/0108159.
[25] H. Hori, S. Sonoda, T. Sasaki, Y. Yamamoto, S. Shimizu, K. Suga, and K. Kindo, submitted for publication to Physica B, cond-mat/0203223.
[26] T. Sasaki, S. Sonoda, Y. Yamamoto, S. Shimizu, K. Suga, K. Kindo and H. Hori, Appl. Phys. Lett., vol.91, 7911, 2002.
[27] M. Sato, H. Tanida, T. Sasaki, Y. Yamamoto, S. Sonoda, S. Shimizu, and H. Hori, to be published Jpn. J. Appl. Phys.
[28] M. Kamp, M. Mayer, A. Pelzmann, K. J. Ebeling, Internet Journal of Nitride Semiconductor Research, vol.2, 26, 1997.
[29] M. Katayama, E. Nomura, N. Kanekama, H. Soejima, and M. Aono, Nucl. Instrum. Methods B 33, pp. 857, 1988.
[30] M. Katayama, R. S. Williams, M. Kato, E. Nomura, and M. Aono, Phys. Rev. Lett., vol. 66, pp. 2762, 1991.
[31] T. Okuno, T. Fujino, S. Shindo, M. Katayama, K. Oura, S. Sonoda and S. Shimizu; Jpn. J. Appl. Phys., vol. 41, L415, 2002.
[32] A. Oiwa, S. Katsumoto, A. Endo, M. Hirasawa, Y. Iye, H. Ohno, F. Matsukura, A.Shen and Y. Sugawara: Solid state Communication 103, pp. 209, 1997.
[33] H. Ohno, SCIENCE, vol.281, 951, 1998.
[34] A. Van Esch, L. Van Bockstal, J. De Boeck, G. Verbanck, A. S. van Steenbergen, P. J. Wellmann, B. Grietens, R. Bogaerts, F. Herlach and G. Borghs; Phys. Rev. B, vol. 56, 13103, 1997.
[35] K. Suga, K. Kindo, T. Sasaki, Y. Yamamoto, S. Sonoda, S. Shimizu and H. Hori, to be submitted to J. Phys. Soc. Jpn.
[36] Y. Yamamoto, H. Hori, T. Sasaki, S. Sonoda, S. Shimizu, K. Suga, and K. Kindo, preparation for publication.